# Particle Swarm Optimization for Realizing Intelligent Routing in Networks with Quality Grading

T. R. Gopalakrishnan Nair
Advanced Networking Research Group
Director, Research and Industry Incubation Centre (RIIC)
Dayananda Sagar College of Engineering
Bangalore, India- 78
e-mail: trgnair@ieee.org

Kavitha Sooda
Advanced Networking Research Group (RIIC), DSI
and Asst. Professor, Dept. of CSE
Nitte Meenakshi Institute Technology
Bangalore, India-64
e-mail: kavithasooda@gmail.com

*Abstract*— Significant research has been carried out in the recent years for generating systems exhibiting intelligence for realizing optimized routing in networks. In this paper, a grade based two-level based node selection method along with Particle Swarm Optimization (PSO) technique is proposed. It assumes that the nodes are intelligent and there exist a knowledge base about the environment in their local memory. There are two levels for approaching the effective route selection process through grading. At the first level, grade based selection is applied and at the second level, the optimum path is explored using PSO. The simulation has been carried out on different topological structures and it is observed that a graded network produces a significant reduction in number of iteration to arrive at the optimal path selection.

## I. INTRODUCTION

The research in the recent years has shown significant interest in bringing out a change in scenario from fixed protocol paradigm to a more adaptive network paradigm for realizing better management and efficiency. By making the network to think, remember and act, we shall be able to achieve more efficient network operations. The aspect of such intelligence has been dealt in cognitive network [1] and autonomic networking [2]. Intelligent network was well formulated based on heuristic algorithm, bio-inspired computing, evolutionary algorithm and human immune system. Applying intelligent methods in networks and gaining higher efficiency in operation have a greater significance in the area of optimal communication, realizing best Quality of service and security, development of application software, and the management of different layers of protocol.

This paper presents the implementation of a novel idea of applying intelligence which enables the nodes to take decisions and carry out the routing of the packets effectively. The results show that the grade routing approach along with PSO has achieved the optimal path with promising nodes. PSO is a method of optimizing the candidate results iteratively and trying to progress towards the final result. Simulation studies show that the optimization method contains few constraints. The resources involved in the implementation of such an algorithm is less and it can be easily used for a real-time environment. The maximization of Quality properties is not seriously considered in today's network scenario for solving optimal path. The introduction of quality grading is found to be helpful in obtaining quality property maximization while generating routing optimization.

The rest of this paper is organized as follows: In Section II the related literature is described and in the next section proposed work is presented in detail. Simulation results and analysis are presented in Section IV. Conclusions of the paper are provided in the last section.

## II. RELATED WORK

Several algorithms have been implemented and analyzed for solving shortest path problem [3]. Although these algorithms are simple and easy to use, they have several drawbacks which make them unsuitable for dynamically changing environment. Here, the routing decisions are merely based on the table information, provided at the node level. Such type has no real awareness of the environment around it. This is where autonomic networks can play its role through the intelligence aspect incorporated in it. Many researchers are working in the area of autonomic network (AN) since a decade.

The ability to be aware of network operation and subsequently adjust the operational parameters according to the needs of the scenario is referred to as cognitive. It behaves based on a reference to an active network but also include an adaptation and learning technique [4, 5] which makes the process different.

Cognition was conceptualized by Mitola [6] and later the idea of feedback loop [7] was derived from it. With advancement in the field of cognition, autonomic approach became more evident to solve a number of network related issues. This approach was also applied to business aspects, technical issues and different functionality at layers. In [2], the author states that in order to become autonomic, it is not sufficient if the system possess the self- properties. Along with self- learning, the system needs to be environment-aware. This awareness adds to the capability of knowledge gain, monitoring and adjustment. It is a challenging task to realize a representation module and a knowledge understanding system in autonomic systems.

## III. PARTICLE SWARM OPTIMIZATION AND GRADED NETWORK

Grade value estimation method for implementing intelligent routing in autonomic network is the core thrust and focus of research. This paper has defined grade as an index which is made available everywhere and routing depends on it. It signifies the quality of the router [8]. Router must be an intelligent node because it has different operation to perform based on the environment information.

### A. Particle Swarm Optimization (PSO)

PSO is an optimization technique developed by Kennedy and Eberhart [10], and is inspired by the social behaviour of bird flock. Input to the PSO algorithm is given in the form of particles, hence in the case of networks, the multiple paths (between two nodes) obtained is encoded as particles. Here we use a modified version of Indirect Encoding Technique. Particles (or sequence of nodes or path) obtained by indirect encoding scheme, serve as input to the PSO algorithm. Here, our objective is to find the particle which has the maximum cost (bandwidth), associated with the links of the particle (or path). In [11] the formulae for velocity and position have been specified. The particle's best position is referred to as *pBest* and the best position in comparison to all other particles is referred to as *gBest*, i.e. the *global best*.

In every iteration, the *pBest* is calculated, using which we obtain the *gBest* value, which is the shortest path of the network. The *pBest* value is calculated depending on the value of fitness of each particle in every iteration. The fitness value depends upon the bandwidth associated with the particles. The fitness value is calculated as below,

$$Fitness = \frac{B_{Initial\ Link}}{\sum_{i=0}^{l_{particle}} B_i} \qquad (1)$$

Here $B_{Initial\ Link}$ represents the bandwidth of the first link of a given particle and $B_i$ represents the total bandwidth of the particle. More the fitness value less the chances of corruption of data while transmission.

### B. Path Encoding for PSO

The nodes of the network are assigned with different priorities, which are used for encoding of the path. Highest priority indicates that the node is active and the data can be sent through it without getting corrupted.

Path construction often leads to formation of loops, hence to avoid this, the selected nodes are assigned a very large negative value as their priority. Here in order to avoid backtracking, a *heuristic operator M* is applied. *M* is considered to be a constant value, 4 as in [11]. But since we are using a region based network, the *M* value has been modified and assigned the *pnr* value. i.e., *M* is now the number of nodes present in each region.

The pseudo code for the path construction is as follows:
**Step1:** Let k=0, $t^k$=source, $vp^k$= {source}, $x^k$=x, $t^k$ = source, $x^k(t^k)$ = -999
**Step2:** If $t^k$ =destination go to step4, else k=k+1 and go to step3.
**Step3:** Select the node with the highest priority from amongst the nodes that are directly linked with the node $t^k$ -1 as $t^k$ if ($t^k$ - $t^{k-1}$)>-M (if source>destination) else ($t^k$ - $t^{k-1}$)<M, $vp^k$ ={ $vp^k$ -1, $t^k$ } and $x^k(t^k)$=-999.
**Step4:** Return valid path $vp^k$ or invalid path if terminal node is not the destination node.

Here $t^k$ is the current or terminal node in the partial path, $vp^k$, is a path under process which contains k+1 nodes. $x^k$ is the priority vector which contains the priority value referred to by x. Assigning -999 to the nodes selected in the partial path avoids selection of the same node again for a given path $vp^k$.

### C. Level-1 and Level-2 operations

*Level-1* is applied region-wise as in Fig. 1 and the goal is to achieve favorable routing based on selected attributes. The values obtained from the *Level-1*, must be able to eliminate the non-production node. Non-production nodes are the once which come in the pitfall of congestion and possess less resource availability. This is identified by assigning a grade value from -3 to +3 as in Fig. 2. It signifies the productivity value of that node. At this point we shall be able to calibrate the routing process region wise. Now the graded function i.e., *Level-2*, considers the output of *level-1* values as its input. The output of this function defines the route availability for the set of nodes under consideration. This shall be calculated for all the available paths leading towards the destination node.

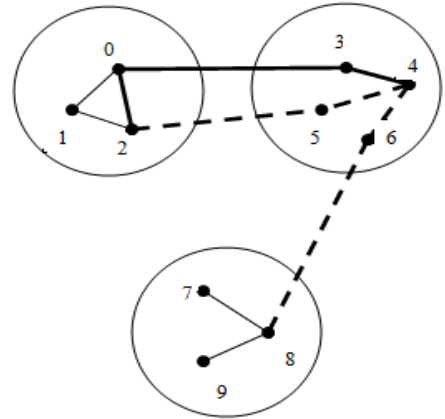

Figure 1. Region Based Network Topology.

*Level-1 operation*
Based on priority assigned top three attributes are selected for every region.
**Step 1:** Top three priority nodes are selected.
**Step 2:** Select the nodes which are nearing to the optimal value (relaxation of ± 2).

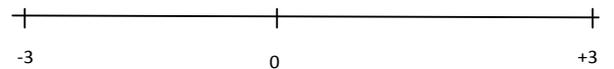

Figure 2. Scale of prioritization assigned to the nodes after observation. Here -3 represents no network lifetime, 0 represents most optimal node and +3 represents nodes which can be considered.

**Step 3:** Selection is done for the nodes which would satisfy the relaxation range from 0 to 2.
**Step 4:** Build a graph to ensure connectivity exists between regions and apply second level grade to find the optimal path.

*Level-2 operation*

The topology obtained from level-1 operation is considered as input data. PSO selection mechanism is applied based on the bandwidth availability for the path determination.
**Step 1:** Consider all possible paths from source to destination as input set.
**Step 2:** Map all the multiple paths from source to destination as particles,
   a. Calculate pBest[] and position of corresponding particle.
   b. for(i=0;i<no_of_iteration;i++)
      for(x=0;x<particle_count;x++)
         Change cost of each link along the path of each particle randomly.
         Calculate fitness and find pBest[] of $i^{th}$ iteration.
   c. Find best pBest[] i.e. gBest & corresponding particle, which gives the shortest path in terms of the cost associated with the links on the path (or particle).

*Methodology and Design*

The design involves the generation of an input model, priority model, gradient algorithm and knowledge base. The input model is based on the M/M/1 technique. The priority model is obtained as shown in Fig. 4. The gradient algorithm is the level-1 operation as discussed earlier. The knowledge base contains the nodes and paths with good condition. Six parameters are considered to obtain the required result. The following parameters are made available in a vector: Resource allocated, Network lifetime and Bandwidth at nodes. *Delay*, *congestion* and *Node Density* are calculated. *Delay* is derived from service rate, arrival rate and capacity. *Congestion* is derived from the expected data rate at the nodes. *Node Density* to be calculated is based on in-degree of the topology been setup. Information about the connectivity must be made available in a file.

Fig. 3 depicts the flow of implementation. Initially the network is studied by input M/M/1 model. In the next stage the priority model selects certain number of nodes by *Level-1 operation*. Later *Level-2 operation* is applied by which we determine the optimal path. This is the knowledge gained by the network and is fed to the knowledge base.

For the first level of node selection we apply the *level-1* operation. The nodes are prioritized based on local observation. Prioritizing is performed based on ontological reasoning as follows:

```
IF (NL) {
    IF (ND < 5) {
        IF (TC does not exist){
            IF (RA){
                IF(Delay does
                not exist){
                    P=1;
                ELSE
                    P=2;
            ELSE
                P=3;
        ELSE
            P=4
    ELSE
        P=5;
ELSE
P=6;
```

Figure 4. Priority Model

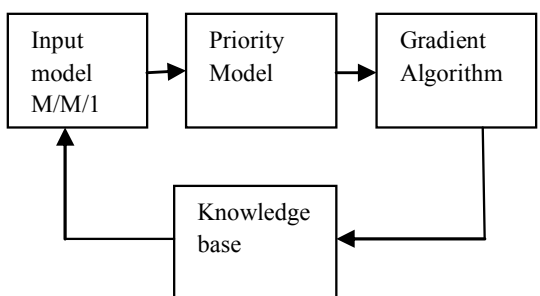
Figure 3. Flow diagram of the implementation

*Delay calculation*

We assume that the node location, external traffic requirements $\gamma_{ik}$, channel cost $d_i(C_i)$, the constants D, µ and the flow ($\lambda_i$) are given and feasible. Thus the Delay T at the node, is given by (2).

$$T = \sum_{i=1}^{M} \frac{\lambda_i}{\gamma} \left[ \frac{1}{\mu C_i - \lambda_i} \right] \quad (2)$$

The average rate of message flow $\lambda_i$, on the $i^{th}$ channel is equal to sum of the average message flow rate of all paths that traverse this channel. The traffic entering the network from the external sources forms a Poisson's process with a mean $\gamma_{jk}$ (messages per second) for those messages originating at node j and destined for node k. Therefore total external traffic entering and leaving the network are considered to be equal to $\lambda_i$.

D. Queueing Systems

A queueing model is used for mathematically analyzing the queueing behavior. We can obtain many steady state performance measures which include: average number in the queue, probability of finding the system in a particular state and queue being full or empty and statistical distribution of number of queues. The model used for the current work is M/M/1 [9]. It stands for Markovian inter-arrival time along with exponential service time with a single server.

*E. Implementation of grade*

The paper is based on Network lifetime, Node density, Traffic congestion, Resource allocation, Delay of the packet arrival and Bandwidth availability parameters which helps in grading according to priority.

## IV. SIMULATION RESULTS AND ANALYSIS

The topology was setup using region based design approach. A random topology was setup and tested for different number of nodes. Initially, the path selection took place region wise. Later the regions were considered for connectivity based on the initial setup. Six parameters are considered for the simulation. Five parameters are considered by *Level-1* and sixth parameter is considered by *Level-2* approach which involves PSO.

TABLE I. COMAPARISON OF FITNESS VALUE FOR PSO AND GRADED PSO

| Iterations | PSO | | Graded PSO | |
|---|---|---|---|---|
| | Iteration | Fitness | Iteration | Fitness |
| 11 | 11 | 0.8 | 10 | 0.8 |
| 12 | 12 | 0.85 | 9 | 0.85 |
| 13 | 13 | 0.80 | 10 | 0.80 |
| 14 | 14 | 0.781818 | 11 | 0.781818 |
| 15 | 15 | 0.8 | 10 | 0.8 |
| 16 | 16 | 0.781818 | 13 | 0.781818 |
| 17 | 17 | 0.8 | 16 | 0.8 |
| 18 | 18 | 0.8 | 17 | 0.8 |
| 19 | 19 | 0.85 | 15 | 0.85 |
| 20 | 20 | 0.8 | 15 | 0.8 |

Table 1, gives a comparison of optimal path determination of PSO with and without quality grading of nodes for one of the random run of the model. In order to meet the size constraint, results of the second half is shown. Here, the number of iteration required to obtain the optimal path was less in graded PSO than without grading PSO technique.

Fig. 5 shows that the number of nodes selected by the PSO with quality grading of nodes was lesser and more reliable than the PSO without quality grading of nodes.

## V. CONCLUSION AND FUTURE WORK

The result obtained shows a significant improvement in the convergence of optimal path by using PSO with quality grading of nodes when compared to a non-graded network. The comparative results of the two approaches show that the graded PSO selected only the best nodes for path determination when compared to the PSO operation carried out on a network having no awareness of the state of the network.

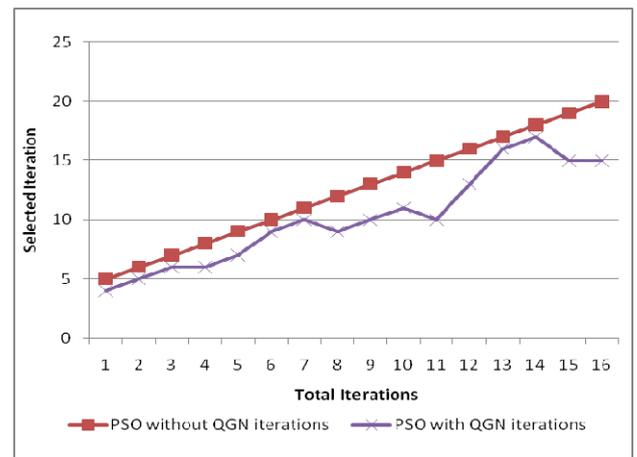

Figure 5. The graph shows the nodes selected by the two methods.

The implementation may further be improved with many more parameters that need to be considered to assess the grade function. These can be applied on a homogeneous or a non-homogenous set of nodes producing grades. This graded network helps in decision making using intelligent arbitration.